\documentclass[a4paper,twocolumn,aps]{revtex4}
\usepackage{graphicx}
\usepackage{amssymb}

\begin{document}
\title{Concentration Dependence of the Flory $\chi$ Parameter
within Two-State Models}
\author{V.A. Baulin and A. Halperin\footnote{To whom
correspondence should be addressed, \\e-mail:
halperin@drfmc.ceng.cea.fr}} \affiliation{UMR 5819 (CEA, CNRS,
UJF), DRFMC/SI3M, CEA-Grenoble, 17 rue des Martyrs, 38054 Grenoble
Cedex 9, France}

\begin{abstract}
The Flory $\chi$ parameter is typically assumed to depend only on
the temperature, $T$. Experimental results often require the
replacement of this $\chi(T)$ by $\chi_{eff}$, that depends also
on the monomer volume fraction, $\phi$, $\chi_{eff}(\phi,T)$. Such
$\chi_{eff}(\phi,T)$ can arise from two state models, proposed for
polyetheleneoxide (PEO) and other neutral water-soluble polymers.\
The predicted $\phi$ dependence of $\overline{\chi
}=\chi_{eff}-(1-\phi)\partial \chi_{eff}/\partial \phi$,
obtainable from colligative properties, differs qualitatively
between the various models: (i) The model of Karlstrom ({\em J.
Phys. Chem.} {\bf 1985}, 89, 4962) yields $\partial \overline{\chi
}/\partial \phi \geq 0$ while the model of Matsuyama and Tanaka
({\em Phys. Rev. Lett.} {\bf 1990}, 65, 341) and of Bekiranov {\em
et al} ({\em Phys. Rev. E} {\bf 1997}, 55, 577) allows for
$\partial \overline{\chi }/\partial \phi <0$ (ii)
$\overline{\chi}(\phi)$ as calculated from the Karlstrom model,
utilizing the parameters used to fit the phase diagram of PEO,
agrees semiquantitatively with the experimental values. On the
other hand, $\overline{\chi}(\phi)$ similarly calculated from the
model of Bekiranov {\em et al }differs qualitatively from the
measured results. Altogether, $\overline{\chi}(\phi)$ provides
useful measure for the performance of a model.
\end{abstract}

\maketitle

\section{Introduction}

The Flory free energy plays a central role in the thermodynamics
of polymers. In this free energy, the mixing energy term has the
form $\chi \phi(1-\phi)$ where $\chi$ is the Flory interaction
parameter and $\phi$ is the monomer volume fraction. Typically, it
is assumed that $\chi$ depends only on the temperature, $\chi
=\chi (T)$.\cite{Flory,PGG,GK,Doi} However when thermodynamic data
are analyzed in terms of the Flory free energy, it is often
necessary to replace $\chi \phi(1-\phi)$ by
$\chi_{eff}\phi(1-\phi)$ where $\chi_{eff}$ is a function of both
$T$ and $\phi$ {\em i.e.}, $\chi_{eff}=\chi
_{eff}(T,\phi)$.\cite{Flory,wolf,Molyneux,Rowlinson,nom}
Considerable effort was devoted to clarifying the statistical
mechanical origins of this and other deviations from the
Flory-Huggins theory. Various extensions of the Flory-Huggins
theory allow for compressibility,\cite{sanchez} local
composition\cite{localcomp} and the role of intrachain
contacts.\cite{Painter} Recently, the lattice cluster theory
utilized a more accurate solution of the lattice model while
allowing for the structure of the monomers.\cite{Freed} The
Flory-Huggins theory was also extended in order to account for the
phase behavior of aqueous solutions of neutral water-soluble
polymers and in particular polyetheleneoxide (PEO). In such
systems the solubility of the polymers is attributed to the
formation of H-bonds. These extensions introduce a {\em two-state
model} wherein the monomers of the chain can exist in two distinct
and interconverting states\cite{Goldstein,K,tanaka,n-Cluster,BBP}
(Figure 1). In the following we demonstrate that ``two-state''
models yield $\chi_{eff}(\phi)$ and analyze the $\phi $ dependence
of the resulting $\chi_{eff}$. Our discussion is mostly concerned
with two models: The one proposed by Karlstrom (K model)\cite{K}
and the model of Matsuyama and Tanaka\cite{tanaka} as formulated
by Bekiranov, Bruinsma and Pincus (MB model).\cite{BBP}

\begin{figure}
\includegraphics[width=1.00\columnwidth,
  keepaspectratio]{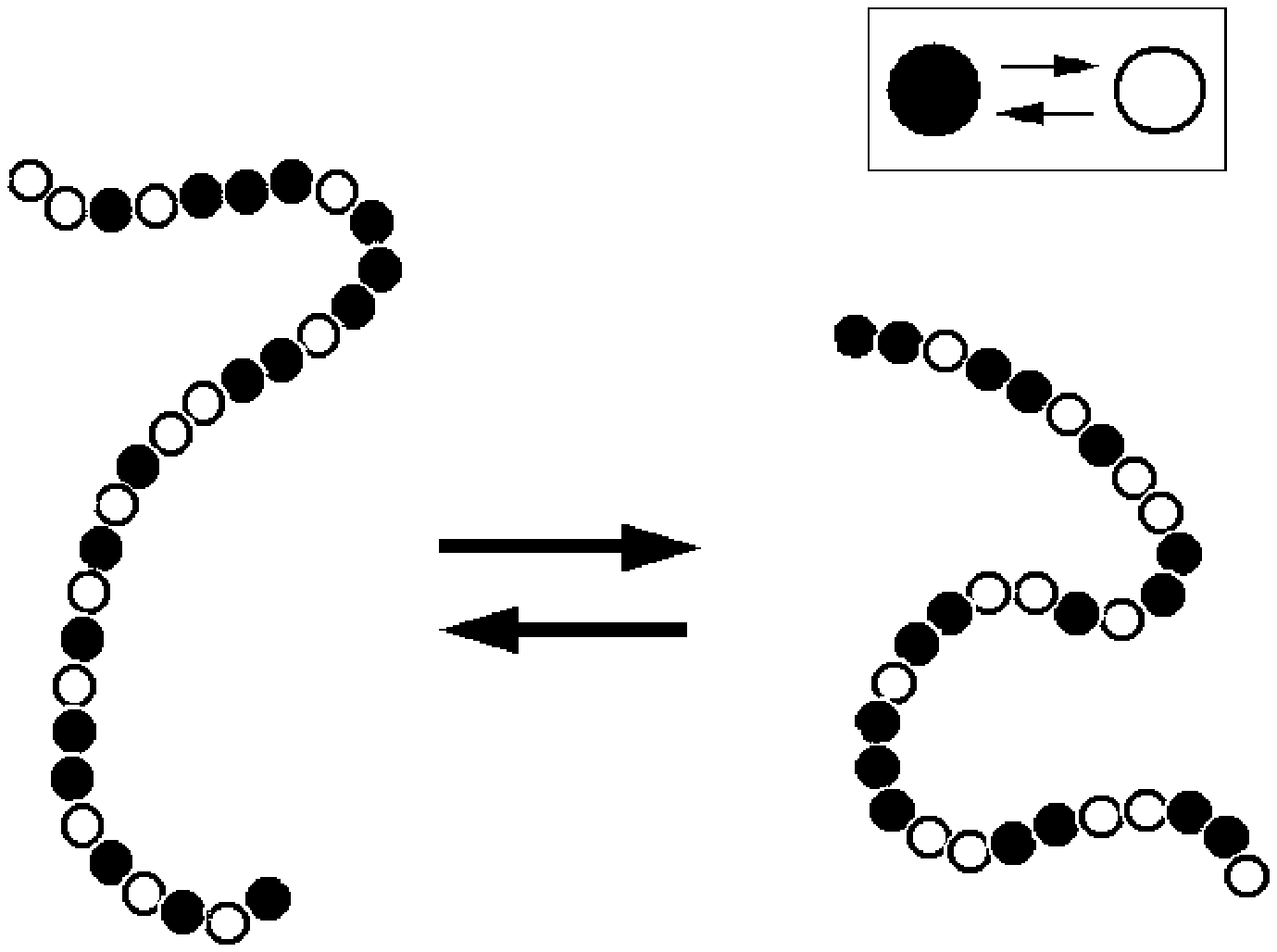}

Figure 1. A schematic representation of the dynamic equilibrium in
two-state polymers where the monomers exist in two interconverting
states (inset), $A$ (empty circle) and $B$ (full circle).
\vspace{-0.5cm}
\end{figure}

The two models differ, as we shall discuss, in the precise
identification of the monomeric states. However, in both cases it
is assumed that one monomeric state, $A$, is ``hydrophilic'' while
the other, $B$, is ``hydrophobic''. As a result, a binary solution
of a polymer in water behaves in fact as a {\em reactive ternary
system}. In particular, the fraction of $A$ monomers, $P$, depends
on both $T$ and $\phi$. Consequently the different versions of
this model yield, as we shall see, $\chi_{eff}$ that depends on
both $T$ and $\phi$. The original papers proposing theses models
\cite{K,tanaka,BBP} focused on the phase behavior while
overlooking the emergence of $\chi_{eff}(\phi)$\cite{foot}.

The goal of our work is to obtain explicit expressions for
$\chi_{eff}(\phi)$ for the K and MB models, analyze the $\phi$
dependence and compare the results to experimental data. It is
noteworthy that by allowing for the internal degrees of freedom of
the monomers, the two-state models can rationalize the observed
$\chi_{eff}(\phi)$ while retaining the main approximations of the
Flory-Huggins theory {\em i.e.}, incompressibility, monomers and
solvent molecules of identical size and shape and local
concentration that equals the global one. Aside from the
fundamental interest in the origin of the $\phi$ dependence of
$\chi_{eff}$ this analysis underlines the utility of
$\chi_{eff}(\phi)$ as a criterion for the performance of the
models.
\begin{figure}
\includegraphics[width=0.80\columnwidth,
  keepaspectratio]{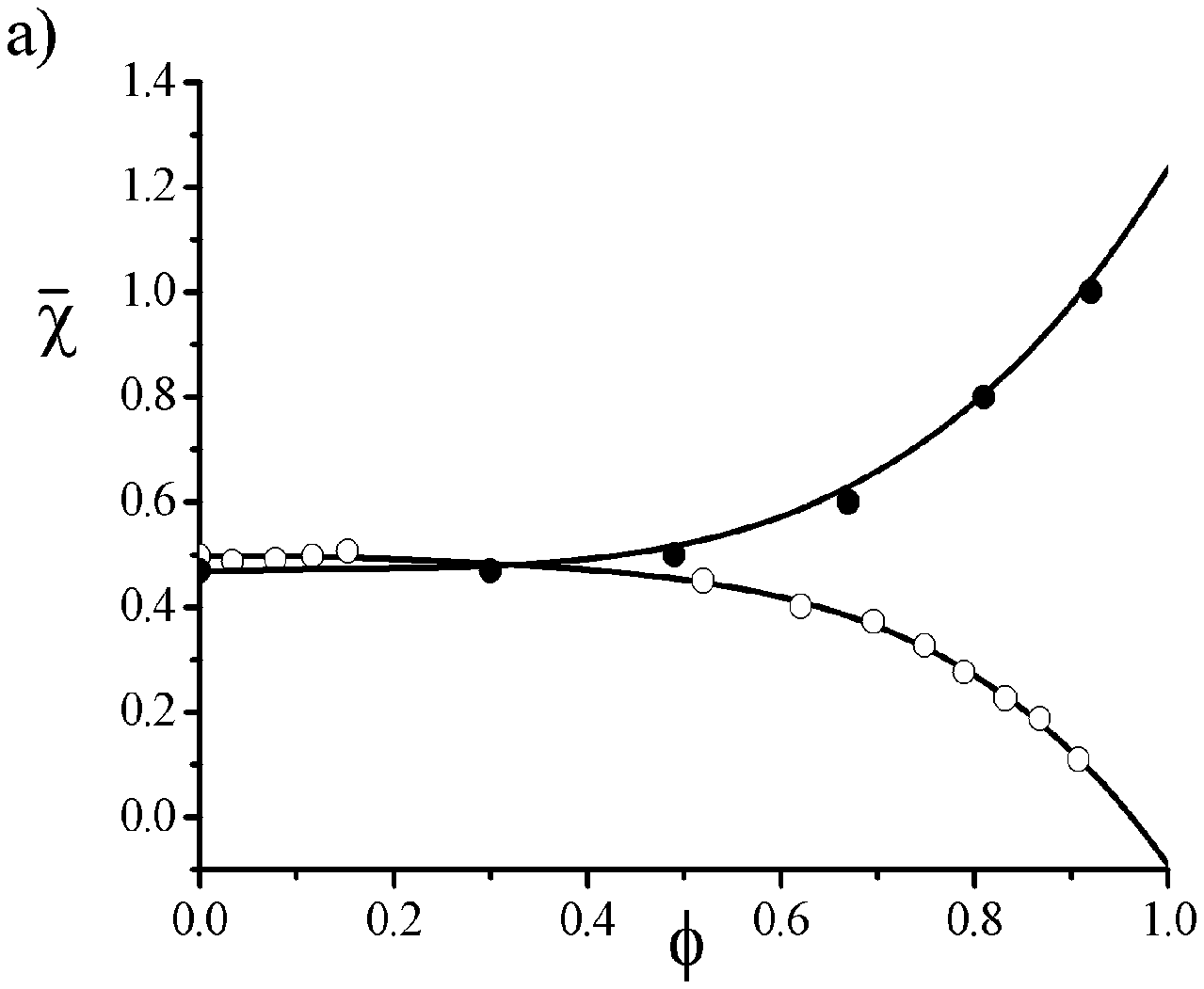}
\end{figure}
\begin{figure}
\includegraphics[width=0.80\columnwidth,
  keepaspectratio]{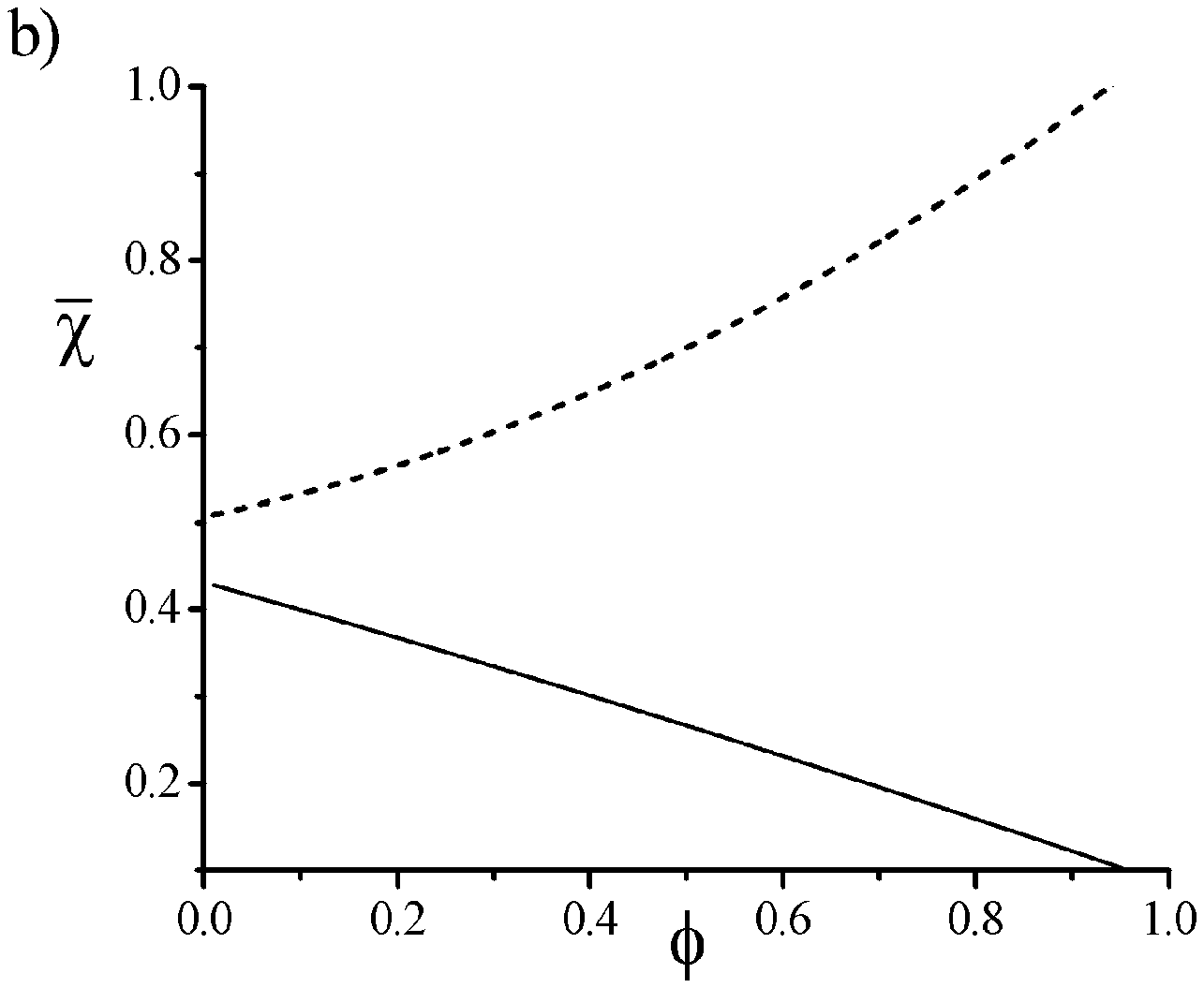}

Figure 2. Plots of measured $\overline{\chi }=\chi_{eff}-(1-\phi
)\partial \chi _{ef}/\partial \phi$ {\em vs}. $\phi$ for (a)
aqueous solutions of the neutral water soluble polymers PEO
$N=3000$, $T=338K$ (full circle, after\cite{Molyneux,Rowlinson})
and PVP $T=295K$ (empty circle, after\cite{Martien}) and for (b)
nonaqueous solutions of polyisobutylene in benzene (dashes) and of
polystyrene in toluene (continuous line)\cite{wolf}.
\vspace{-0.5cm}
\end{figure}

While both the K model and the MB model recover the observed phase
diagram of PEO, only the K model yields a qualitatively correct
$\chi_{eff}(\phi )$. On the other hand, an opposite trend is found
for $\chi_{eff}(\phi)$ of polyvinylpiridon (PVP) which exhibits a
$\phi$ dependence that differs qualitatively from that of PEO
\cite{Martien}(Figure 2). In this case the K model is inherently
incapable of reproducing the experimental results while the MB
model does allow for a similar behavior.

Our analysis is mostly concerned with the concentration dependence of
\begin{equation}
\overline{\chi }=\chi _{eff}-(1-\phi )\frac{\partial
\chi_{eff}}{\partial \phi}.  \label{I1}
\end{equation}
We choose to focus on $\overline{\chi}$ because measurements of
colligative properties, such as vapor pressure and osmotic
pressure, yield $\overline{\chi}$ rather than $\chi
_{eff}$.\cite{wolf} As a result, it is the $\overline{\chi}$
values that are usually reported in the literature. Furthermore,
it turns out that the $\phi$ dependence of $\overline{\chi}(\phi)$
is simpler to analyze. Since the different versions of the
two-state model aimed to rationalize the phase behavior of aqueous
solutions of PEO, the parameters involved were chosen with view of
recovering the known phase diagram. However, these models are in
fact suitable candidates for the description of aqueous solutions
of neutral water-soluble polymers in general. With this in mind we
consider the two-state model as a generic model for such systems
rather than as a specific model for PEO. Accordingly, we make no
specific assumptions regarding the choice of the various
parameters involved.

The paper is organized as follows. A unified description of the K
and MB models, stressing the similarities and the differences, is
presented in section II. Section III contains a brief summary of
the thermodynamics of polymer solutions with $\chi_{eff}(\phi)$.
This discussion introduces the approach utilized later for the
study of $\overline{\chi }(\phi )$. $\chi _{eff}(\phi )$,
$\overline{\chi}(\phi )$ and $\partial \overline{\chi } (\phi
)/\partial \phi$ for the K and MB models are discussed in sections
IV and V respectively. In the discussion we compare the K and MB
models with two other physically transparent models yielding a
$\phi $ dependent $\chi_{eff}$.

\section{Two ``two-state'' Models}

In the ``two-state'' models for water-soluble homopolymers the
monomers of the chain may assume two interconverting forms
characterized by different interaction energies. In particular,
one monomeric state, $A$, is assumed to be ``hydrophilic'' while
the other, $B$, is assumed to be ``hydrophobic''. The K and MB
models differ somewhat with respect to the underlying physics. The
K model\cite{K} focuses on molecular rotations giving rise to
monomeric states of different polarities. The state endowed with a
strong dipole moment is identified as the hydrophilic species,
$A$, while the less polar state is identified as the hydrophobic
state, $B$. Within this model the interconversion is a
unimolecular reaction, $A\rightleftarrows B$, and the mass action
law is
\begin{equation}
\frac{P}{1-P}=K_{K}  \label{II1}
\end{equation}
Here $P$ is the fraction of monomers in a hydrophilic $A$ state
and $K_{K}$ is the corresponding equilibrium constant. The
explicit form of $K_{K}$, to be given in section IV, is irrelevant
at this point. The MB model\cite{BBP} focuses on the role of
H-bonds. In particular, a monomer that forms H-bond with water is
considered, in effect, as hydrophilic while a monomer that does
not form such bond is hydrophobic. Within this model one
distinguishes between free water molecules ($S$) and water
molecules that are bound to the polymer chain. This distinction
has no counterpart in the K model. Accordingly the interconversion
reaction is a bimolecular reaction, $B+S\rightleftarrows A$, and
the chemical equilibrium is specified by
\begin{equation}
\frac{P}{(1-P)\phi _{0}}=K_{MB}  \label{II2}
\end{equation}
where $\phi _{0}$ is the volume fraction of water. The explicit
expression for $K_{MB}$, to be given in section V, is not required
for the present discussion. Thus, in the K model
\begin{equation}
\phi _{0}=1-\phi  \label{II3}
\end{equation}
while in the MB model

\begin{equation}
\phi_{0}=1-\phi-P\phi  \label{II4}
\end{equation}
where $P$ is the fraction of monomers that form H-bonds and
$P\phi$ is the volume fraction of bound water. As a result, the
two models involve different reference states. In particular, the
melt state, $\phi =1$, in the K model is characterized by
\begin{equation}
P(\phi =1)=P_{\ast }\geq 0  \label{II5}
\end{equation}
while in the MB model
\begin{equation}
P(\phi =1)=0  \label{II6}
\end{equation}
Both models utilize a Flory type approach to obtain the thermodynamics of
the solution. In particular, random mixing is assumed and the energy per
site of a polymer-water solution inscribed on a lattice is written as
\begin{eqnarray}
&&\frac{E}{kT} =\phi \left[ P\frac{z\epsilon _{AA}}{2kT}+\left(
1-P\right) \frac{z\epsilon _{BB}}{2kT}\right] +(1-\phi
)\frac{z\epsilon _{SS}}{2kT}+
\nonumber \\
&&\phi (1-\phi)\left[P\chi_{AS}+(1-P)\chi_{BS}\right] +\phi
^{2}\chi_{AB}P(1-P),
\end{eqnarray}
where $z$ is the coordination number of the lattice and $k$ is the
Boltzmann constant. $\epsilon_{AA}$, $\epsilon_{BB}$ and
$\epsilon_{SS}$ denote respectively the energies of $A$, $B$ and
solvent molecules in a pure bulk phase. $\chi_{AB}$, $\chi_{AS}$
and $\chi_{BS}$ denote the Flory interaction parameters
corresponding respectively to $AB,$ $AS$ and $BS$ interactions
(where $\chi_{AS}=\epsilon_{AS}-(\epsilon_{AA}+\epsilon_{SS})/2$
{\em etc}). The first term allows for the energy of the $A$ and
$B$ in a hypothetical pure bulk phase while the second reflects
the energy of the solvent molecules in a pure bulk state. The
interactions between the two monomeric states and the solvent
gives rise to the third term while the last term allows for the
$AB$ interactions. For future reference it is convenient to define
a dimensionless parameter
\begin{equation}
\Delta \epsilon =z\frac{\epsilon_{AA}-\epsilon_{BB}}{2kT}
\label{II8}
\end{equation}
characterizing the difference between the energies of the two
monomeric states. This expression reduces to the familiar Flory
form when $P=1$ or $P=0$ that is, when the monomers have only one
state. The energy $E$ is supplemented by two entropy contributions
per lattice site. One term allows for the different possible
sequences of $A$ and $B$ states along the chain
\begin{equation}
\frac{S_{AB}}{k}=-\phi \left[ P\ln P+(1-P)\ln (1-P)\right]  \label{II9}
\end{equation}
while the second reflects the translational entropy of the chains
\begin{equation}
\frac{S_{trans}}{k}=-\frac{\phi}{N}\ln \frac{\phi}{N}-\phi _{0}\ln
\phi_{0}.  \label{II10}
\end{equation}
The $S_{AB}$ term is distinctive to the two-state models. It
vanishes when the monomers exist in a single state i.e., $P=1$ or
$P=0$. As expected, $S_{trans}$ is identical in form to the
entropy term in the Flory free energy. Altogether, the free energy
of the solution is $F=E-T(S_{AB}+S_{trans})$ and the mixing free
energy is
\begin{equation}
F_{mix}=F(\phi,P)-\phi F(\phi=1)-(1-\phi)F(\phi=0) \label{II11}
\end{equation}

Both models neglect the fluctuations in the number of $A$ monomers
per chain, $m$. Strictly speaking, chains with different $m$ are
distinguishable and should be treated as different chemical
species.\cite{tanaka} Rather than allow for this multiplies
equilibrium the K and MB models consider a single polymer species
characterized by an average $\overline{m}=PN$ where $N$ is the
polymerization degree. This approximation is justified because the
$m\,$\ values are described by a sharply peaked Gaussian
distribution.\cite{BBP} While the two models lead to similar free
energies, the free energies invoked in the K and in the MB-models
differ in two respects. In the MB model (i) The translational
entropy is $P$ dependent because of the $\phi _{0}\ln \phi_{0}$
term and, (ii) since $P(\phi =1)=0$, $F_{mix}$ does not reflect
contributions due to $P(\phi =1)=P_{\ast }\geq 0$ that appears in
the K model.

Finally, it is helpful to note certain additional differences
between the models as described in the original
papers.\cite{K,BBP} While the physical content of these is minor
in comparison to the differences discussed above, they are of
interest for comparison purposes. Thus, the analysis of Bekiranov
{\em et al} focused on the particular case of $\chi _{AB}=0$,
$\chi _{AS}=\chi _{BS}=\chi =2.885-0.0036T$ and $\Delta \epsilon
=6.38-2408/T$ while in the K model $\chi _{AS}=80.0/T,\chi
_{BS}=684.5/T$, $\chi _{AB}=155.6/T$ and $\Delta \epsilon
=-625.2/T+\ln 8$. The temperature dependence of the $\chi $
parameters in the two papers is somewhat different. Bekiranov {\em
et al }assume $\chi \sim a+bT$ while in the K model $\chi \sim
b/T$. In both models $\Delta \epsilon \sim a+b/T$. In the model of
Bekiranov {\em et al} this follows from the identification of
$\Delta \epsilon $ with $\Delta F_{0}/kT$ where $\Delta
F_{0}=\Delta E_{0}-T\Delta S_{0}$ is the free energy change
associated with the formation of H-bond. The $a+b/T$ dependence in
the K model is a consequence of the degeneracy of the monomeric
states. It assumed that each of the two monomeric forms can exist
in a number of equivalent states and the resulting entropy gives
rise to $a$. The analysis of Bekiranov {\em et al} allows for the
effect of pressure. This is attributed to the existence of a
preferred volume for a H-bond.\cite{Poole} The application of
pressure is assumed to introduces a geometric constraint resulting
in a lower number of bound water molecules. This argument yields a
semiquantitative agreement with experimental results on the phase
diagram of PEO. Note that this coupling mechanism can only operate
within the MB model and can not be used in the K model.
Furthermore, it is not fully consistent with the incompressibility
assumption invoked by Bekiranov {\em et al}.

\section{Thermodynamics of Polymer Solutions with $\chi_{eff}(\phi)$}

Before we proceed to analyze the $\phi$ dependence of
$\overline{\chi}$ within the two ``two-state'' models, it is
helpful to summarize the thermodynamic relationships that apply
when $\chi =\chi (T)$ is replaced by $\chi_{eff}=\chi
_{eff}(T,\phi )$\cite{wolf}. The Flory like mixing free energy per
site, $F_{mix}$, consists then of two terms. One is an interaction
free energy that is the counterpart of the mixing energy $\chi
\phi (1-\phi )$
\begin{equation}
\frac{F_{int}}{kT}=\chi _{eff}(\phi )\phi (1-\phi).  \label{III1}
\end{equation}
The second is the translational free energy
\begin{equation}
\frac{F_{trans}}{kT}=\frac{1}{N}\phi \ln \phi +(1-\phi )\ln
(1-\phi). \label{III2}
\end{equation}
The chemical potential of the solvent is $\mu _{s}=\mu
_{s}^{o}(P,T)-\pi a^{3}$ while the osmotic pressure $\pi
a^{3}=\phi ^{2}\frac{\partial}{
\partial \phi }(\frac{F_{mix}}{\phi })$ is
\begin{equation}
\frac{\pi a^{3}}{kT}=\frac{\phi }{N}-\phi -\ln (1-\phi )-\overline{\chi }
\phi ^{2}  \label{III4}
\end{equation}
where $\overline{\chi }=\chi_{eff}-(1-\phi )\frac{\partial \chi
_{eff}}{
\partial \phi }$ replaces $\chi$. Since $\mu_{s}$ determines the
colligative properties of the solution, measurements of such
properties yield $\overline{\chi}$ rather than $\chi_{eff}$. In
turn, $\overline{\chi}$ is obtainable from $F_{int}$ via
\begin{equation}
\overline{\chi }=-\frac{\partial }{\partial \phi
}\frac{F_{int}}{kT\phi} \label{III5}
\end{equation}

In the next two sections we will utilize equation (\ref{III5}) to
obtain $\overline{\chi}$ from $F_{int}=F_{mix}-F_{trans}$. This
equation also shows that terms of the form $const^{\prime }\phi$
in $F_{int}$ do not contribute to $\overline{\chi}$. This will
allow us to ignore such linear terms that arise because of the
choice of the reference state.

\section{$\overline{\chi}$ within the K model}

Within the K model the mixing energy per lattice site, $E_{mix}$,
is

\begin{eqnarray}
\frac{E_{mix}}{kT}&=&\phi P\Delta \epsilon +\phi (1-\phi)[P\chi
_{AS}+(1-P)\chi_{BS}]+ \nonumber \\
&&\phi^2\chi_{AB}P(1-P)+\text{terms linear in }\phi
\end{eqnarray}

\begin{figure}
\includegraphics[width=0.80\columnwidth,
  keepaspectratio]{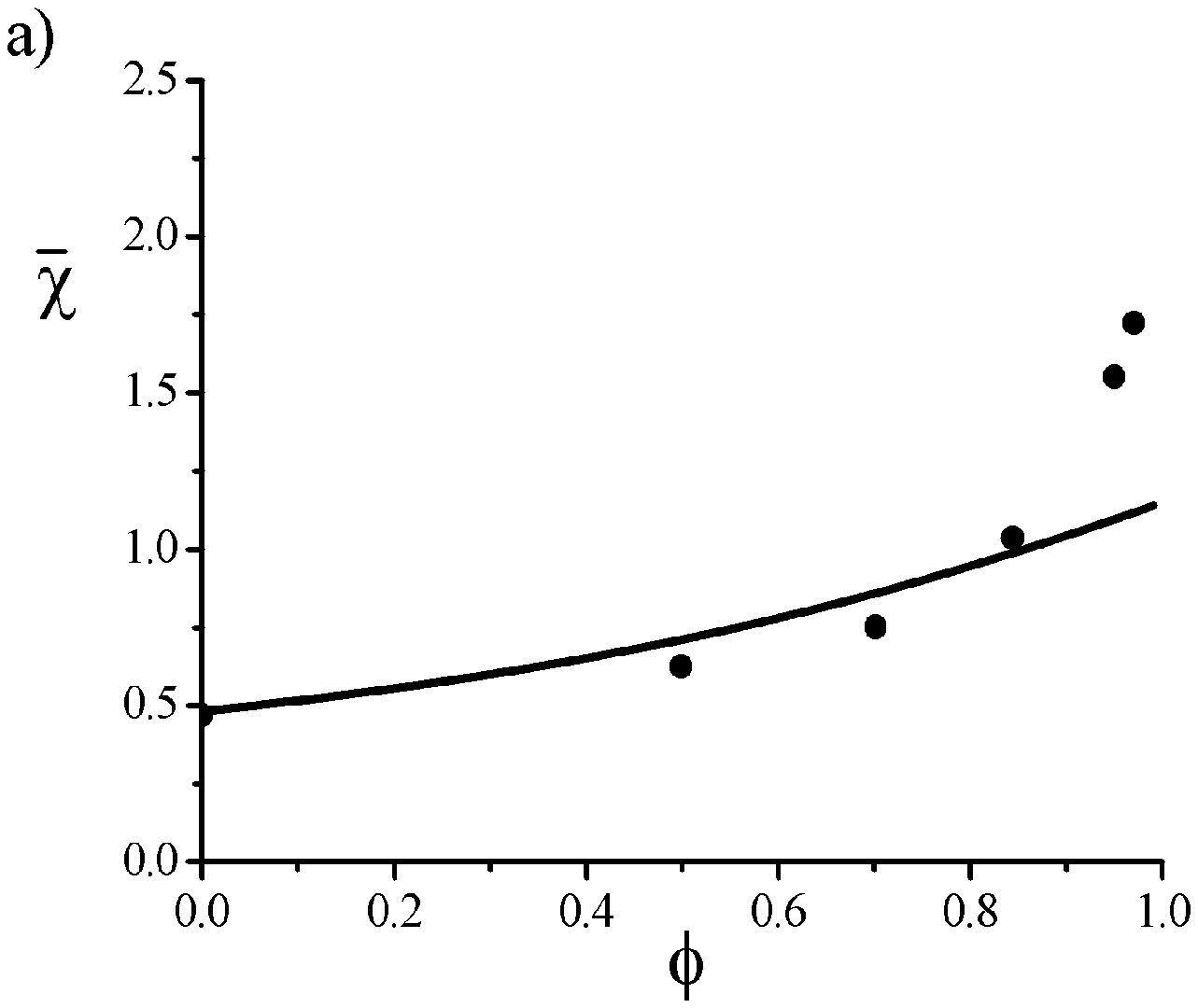}
\end{figure}
\begin{figure}
\includegraphics[width=0.80\columnwidth,
  keepaspectratio]{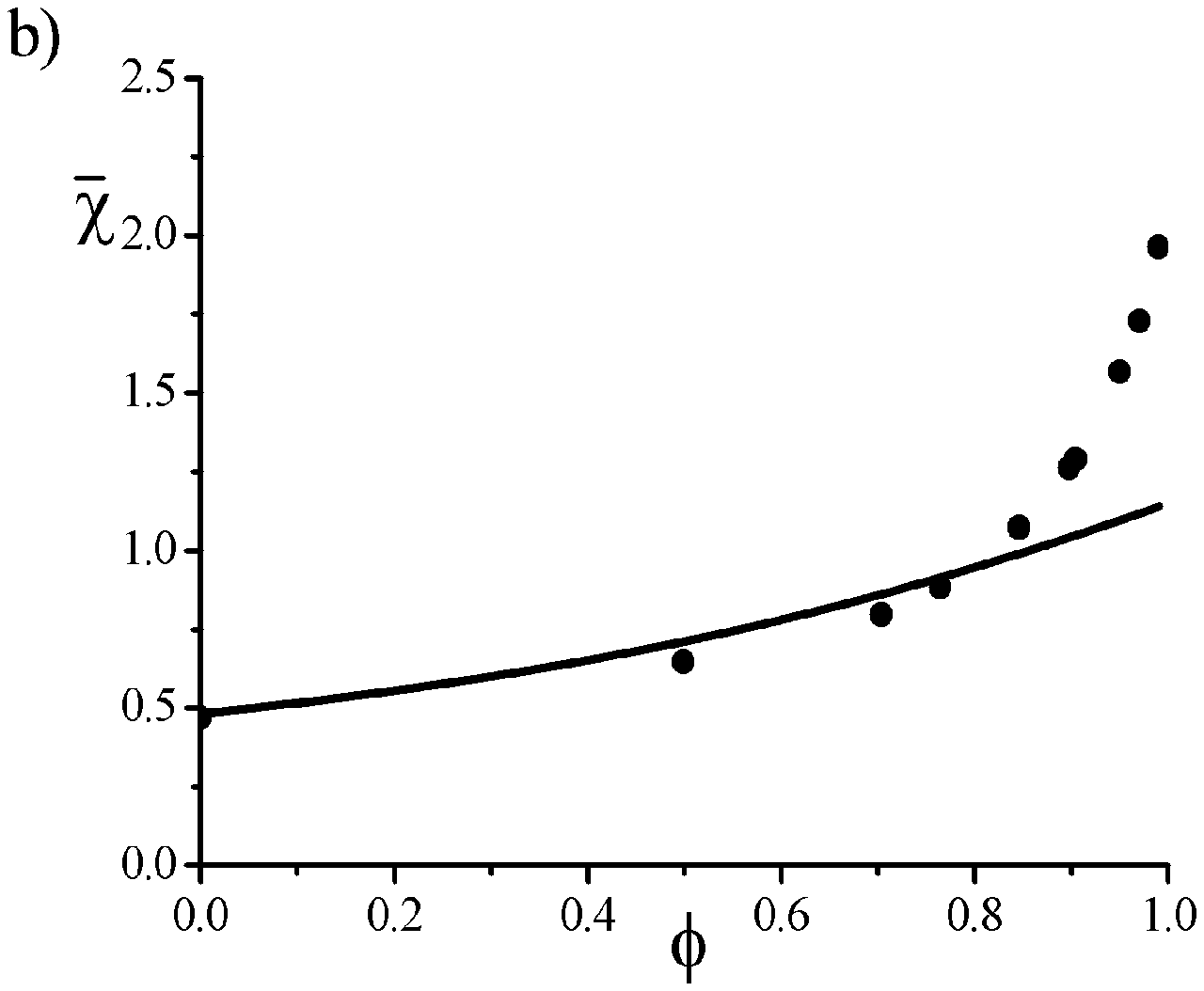}

Figure 3. $\overline{\chi}_{K}$ {\em vs. }$\phi$ (continuous line)
calculated using $\chi _{AS}=80.0/T$, $\chi_{BS}=684.5/T$,
$\chi_{AB}=155.6/T$ and $\Delta \epsilon =-625.2/T+\ln 8$, the
parameters used to fit the phase diagram of PEO, agree
semiquantitatively with the experimental values of $\overline{\chi
}$ (full circles) as measured with $N=3000$ at (a) $T=328K$ and
(b) $T=338K$. While the numerical values and the curvature of
$\overline{\chi}_{K}$ change with the choice of $\chi_{AS}$,
$\chi_{BS}$, $\chi_{AB}$ and $\Delta \epsilon$, it is normally a
monotonically increasing function of $\phi$.
\end{figure}

The first term measures the difference in the energy of isolated
$A$ and $B$ monomers. The second term accounts for the
interactions between the monomers and the solvent. It is a
generalization of the\ $\chi \phi (1-\phi)$ term in the Flory free
energy for a binary solutions. The third term allows for the
interaction between $A$ and $B$ monomers. The mixing entropy is
\begin{eqnarray}
&&\frac{S_{mix}}{k}=-\phi \lbrack P\ln P+(1-P)\ln (1-P)]-
 \\
&&\left[ \frac{\phi }{N} \ln \phi +(1-\phi )\ln (1-\phi)\right]
+\text{terms linear in }\phi \nonumber
\end{eqnarray}
The first term allows for the $AB$ mixing within the chains and the second
for the translational entropy of the chains. In both expressions, the terms
linear in $\phi $ arises because the free energy of the melt state depends
on $P_{\ast }$ and $F_{mix}=F-(1-\phi )F(\phi =0)-\phi F(\phi =1,P_{\ast })$%
.\ Upon discarding terms linear in $\phi $, $F_{int}=F_{mix}-F_{trans}$ is
\begin{eqnarray}
&&\frac{F_{int}}{kT} =\phi P\Delta \epsilon +\phi (1-\phi)[P\chi
_{AS}+(1-P)\chi _{BS}]+  \nonumber \\
&&\phi^2 \chi_{AB}P(1-P)+\phi \lbrack P\ln P+(1-P)\ln (1-P)]
\end{eqnarray}
The equilibrium value of $P$ for a given $\phi$ and $T$,
$P_{e}=P_{e}(\phi)$, is specified by the condition $\partial
F_{int}/\partial P=0$ leading to
\begin{eqnarray}
&&\frac{P_{e}}{1-P_{e}} =K_{K}(\phi ,P_{e},T)=  \label{IV4} \\
&&\exp \left[ -\Delta \epsilon -(1-\phi )(\chi
_{AS}-\chi_{BS})-\phi \chi _{AB}(1-2P_{e})\right] \nonumber
\end{eqnarray}
where $K_{K}$ is the equilibrium constant for the unimolecular
intrachain $A\rightleftarrows B$ reaction. When
$\chi_{AS}=\chi_{BS}=\chi$ and $\chi_{AB}=0$, the parameters
chosen by Bekiranov {\em et. al.}, $K_{K}$ is independent of
$\phi$. Consequently, for this case $P_{e}$ is independent of
$\phi$, $\partial P_{e}/\partial \phi =0$, thus leading, as we
shall see, to $\partial \overline{\chi }_{K}/\partial \phi =0$. To
assure the stability of the $A\rightleftarrows B$ equilibrium we
further invoke the condition $\partial^{2}F_{int}/\partial
P^{2}>0$
\begin{equation}
\frac{\partial ^{2}F_{int}}{\partial P^{2}}=\phi \left[
\frac{1}{P(1-P)}-2\phi \chi _{AB}\right] >0.  \label{IV5}
\end{equation}

This ensures that the free energy curve is concave and that
fluctuation relax back to the equilibrium state according to the
LeChatelier principle. Using $\ \overline{\chi}=-\frac{\partial
}{\partial \phi }\frac{F_{int}}{kT\phi}$ and the equilibrium
condition we find
\begin{equation}
\overline{\chi }_{K}=P_{e}\chi _{AS}+(1-P_{e})\chi_{BS}-\chi
_{AB}P_{e}(1-P_{e})  \label{IV6}
\end{equation}

Thus, within the K model the $\phi$ dependence of $\overline{\chi
}_{K}$ is due to $P_{e}(\phi )$. In order to specify $\partial
\overline{\chi }_{K}/\partial \phi$ it is first necessary to
obtain $\partial P_{e}(\phi )/\partial \phi$ by differentiating
the equilibrium condition (\ref{IV4}) with respect to $\phi$
leading to
\begin{equation}
\frac{\partial P_{e}(\phi )}{\partial \phi }=\frac{\partial
\overline{\chi}_{K}}{\partial P_{e}}\frac{\phi }{\partial
^{2}F_{int}/\partial P^{2}}. \label{IV7}
\end{equation}

Here, and later in (\ref{IV8}), $\partial ^{2}F_{int}/\partial
P^{2}$ signifies the value of (\ref{IV5}) in equilibrium i.e.,
$P=P_{e}(\phi )$. Equation (\ref{IV7}) indicates that the sign of
$\partial P_{e}(\phi )/\partial \phi$ is determined by $\partial
\overline{\chi }_{K}/\partial P_{e}$. This reflects the
concentration dependence of $K_{K}=const^{\prime }\exp(\phi
\partial \overline{\chi }_{K}/\partial P_{e})$. Altogether
\begin{equation}
\frac{\partial \overline{\chi }_{K}}{\partial \phi
}=\frac{\partial \overline{\chi }_{K}}{\partial
P_{e}}\frac{\partial P_{e}}{\partial \phi } =\left( \frac{\partial
\overline{\chi }_{K}}{\partial P_{e}}\right) ^{2} \frac{\phi
}{\partial ^{2}F_{int}/\partial P^{2}}\geq 0  \label{IV8}
\end{equation}
and $\overline{\chi}_{K}$ is an increasing function $\phi$ for any
choice of the parameters $\chi_{AS}$, $\chi_{BS}$, $\chi _{AB}$
and $\Delta \epsilon$ except for the case $\chi _{AS}=\chi_{BS}$
and $\chi_{AB}=0$ when $\partial \overline{\chi}_{K}/\partial
\phi=0$. The calculated $\overline{\chi }_{K}(\phi )$, utilizing
the parameters used to fit the phase diagram of PEO, is in
semiquantitative agreement with the experimental results for
$\overline{\chi}(\phi)$ (Figure 3). The resulting curve is
representative of the behavior of $\overline{\chi}_{K}(\phi)$ in
general.

\section{$\overline{\protect\chi }$ within the MB model}

The mixing energy per site in the MB model is essentially
identical to that of the K model

\begin{figure}
\includegraphics[width=0.80\columnwidth,
  keepaspectratio]{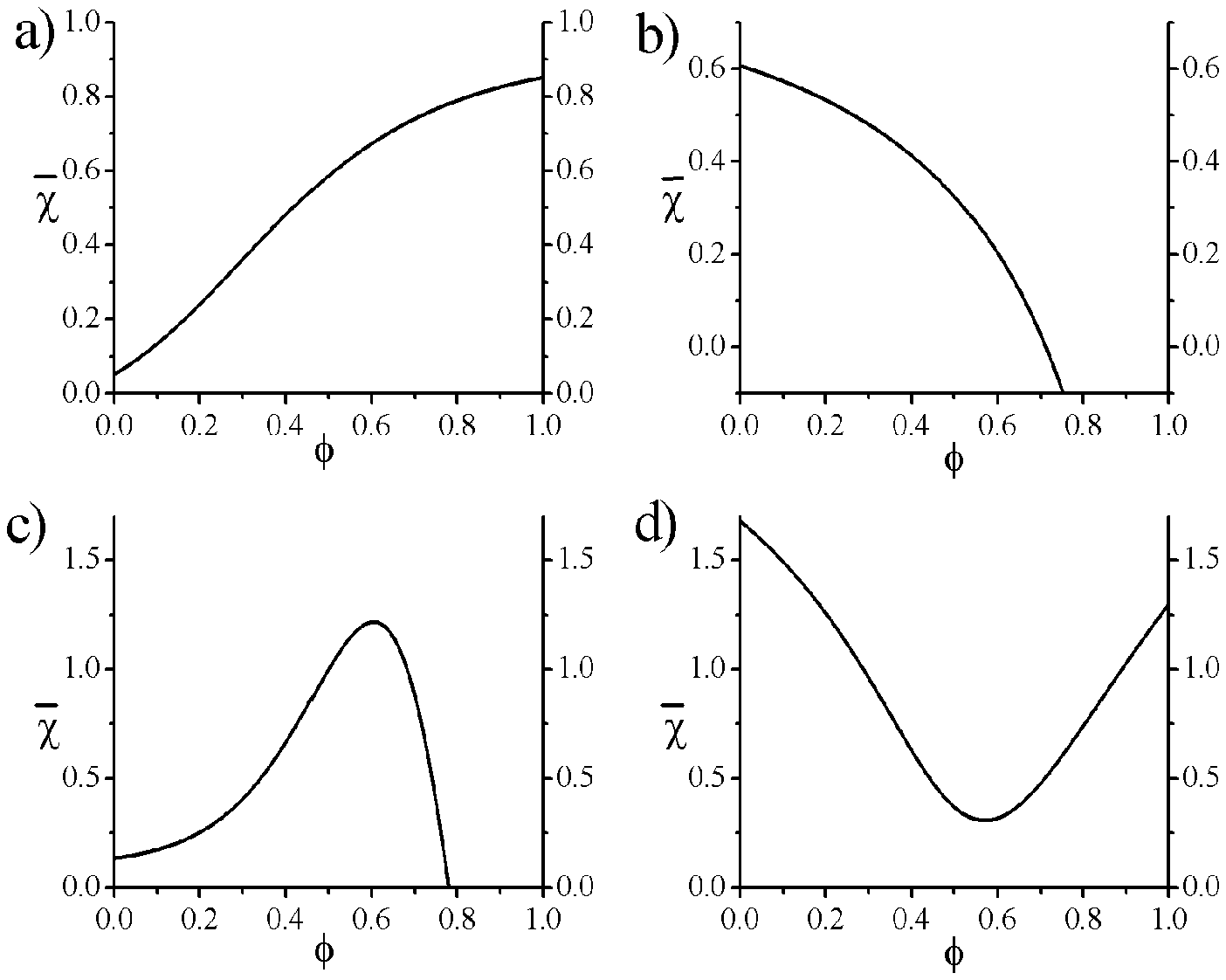}

Figure 4. $\overline{\chi}_{MB}$ {\em vs}. $\phi$ plots exhibit
qualitatively different forms, depending on the choice of
parameters (a) monotonically increasing $\chi_{AS}=0.7$, $\chi
_{BS}=0.9$, $\chi_{AB}=2$ and $\Delta \epsilon=4$, (b)
monotonically decreasing $\chi_{AS}=0.1$, $\chi_{BS}=0.4$,
$\chi_{AB}=-3$ and $\Delta \epsilon=5$, (c) maximum $\chi
_{AS}=9$, $\chi_{BS}=0.1$, $\chi_{AB}=0.1$ and $\Delta \epsilon
=-5$, (d) minimum $\chi_{AS}=3$, $\chi_{BS}=3$, $\chi_{AB}=0.5$
and $\Delta \epsilon =-2$. \vspace{-0.5cm}
\end{figure}

\begin{eqnarray}
\frac{E_{mix}}{kT}&=&\phi P\Delta \epsilon +(1-\phi)\phi
\left[P\chi_{AS}+(1-P)\chi _{BS}\right]+ \nonumber \\
&&\phi^2\chi_{AB}P(1-P) \label{V1}
\end{eqnarray}
while the mixing entropy per site is
\begin{eqnarray}
\frac{S_{mix}}{k}&=&-\phi \lbrack P\ln P+(1-P)\ln (1-P)]-\nonumber \\
&&\left[ \frac{\phi}{N}\ln \phi +(1-\phi-P\phi)\ln (1-\phi
-P\phi)\right]. \label{V2}
\end{eqnarray}

These differ from the corresponding expressions within the K model
in two respects: (i) Terms linear in $\phi$ are no longer present
because the dependence on $P_{\ast}$ disappears since $P(\phi
=1)=0$ (ii) The translational entropy contribution to $S_{mix}$
now depends on $P$ because $\phi _{0}=1-\phi -P\phi$.
$F_{int}=F_{mix}-F_{trans}$ is
\begin{eqnarray}
&&\frac{F_{int}}{kT} =\phi P\Delta \epsilon +\phi (1-\phi)[P\chi
_{AS}+(1-P)\chi_{BS}]+ \nonumber \\
&&\phi^2 \chi_{AB}P(1-P)+\phi \lbrack P\ln P+(1-P)\ln (1-P)]- \nonumber \\
&&P\phi \ln (1-\phi )+(1-P\phi-\phi)\ln \left(1-\frac{P\phi
}{1-\phi}\right).\label{V3}
\end{eqnarray}

The equilibrium value of $P$, for a given $\phi $ and $T$ is specified by
the condition $\partial F_{int}/\partial P=0$
\begin{eqnarray}
&&K_{MB}(\phi,P_{e},T) =\frac{P_{e}}{(1-P_{e})(1-P_{e}\phi -\phi
)}= \label{V4} \\
&&\exp \left[1-\Delta \epsilon -(1-\phi )(\chi _{AS}-\chi
_{BS})-\phi \chi _{AB}(1-2P_{e})\right]  \nonumber
\end{eqnarray}
where $K_{MB}=eK_{K}$ is the equilibrium constant for the
$B+S\rightleftarrows A$ reaction. As in the K model, $K_{MB}$ is
independent of $\phi $ when $\chi_{AS}=\chi_{BS}$ and $\chi
_{AB}=0$.

\begin{figure}
\includegraphics[width=0.80\columnwidth,
  keepaspectratio]{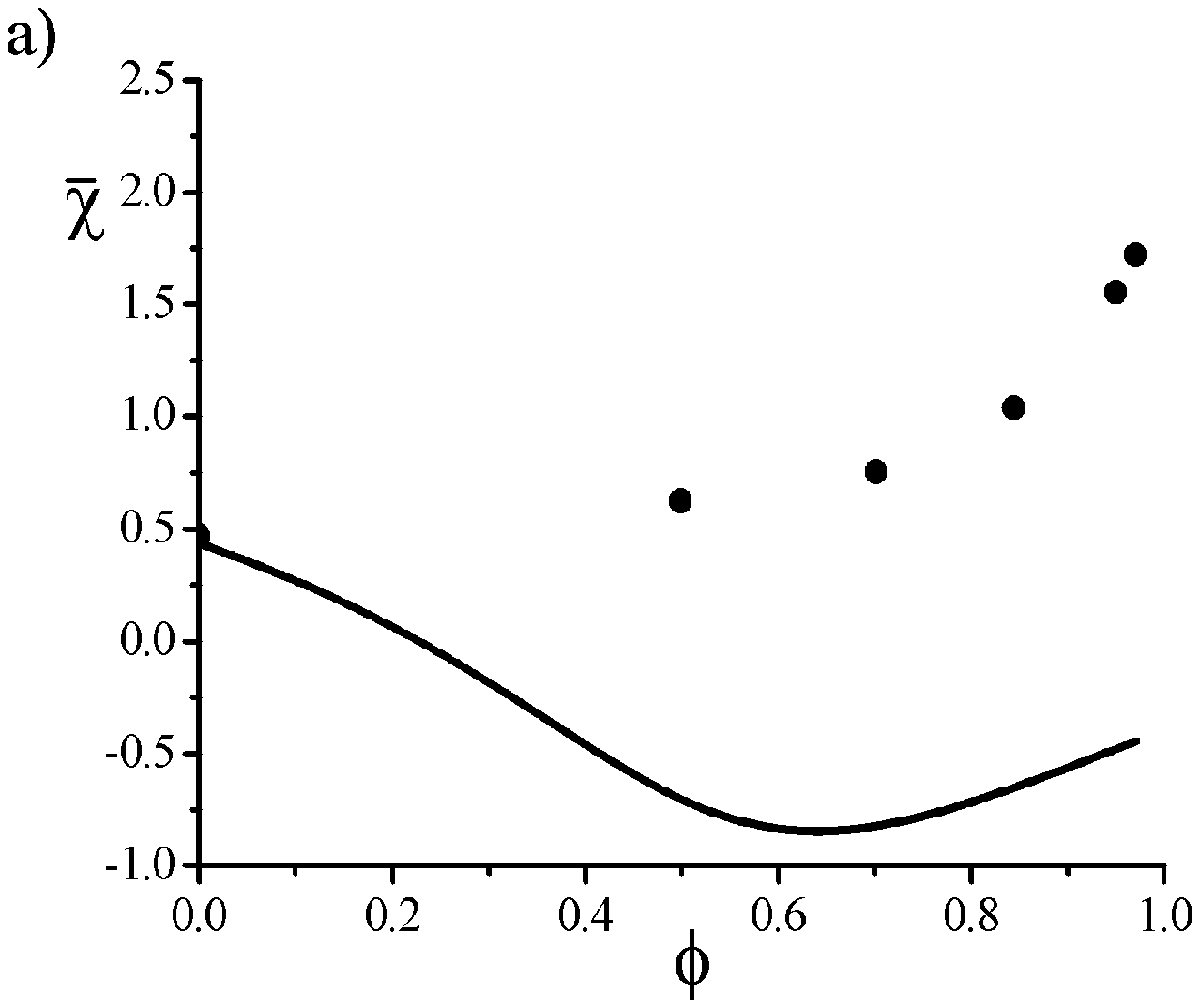}
\end{figure}
\begin{figure}
\includegraphics[width=0.80\columnwidth,
  keepaspectratio]{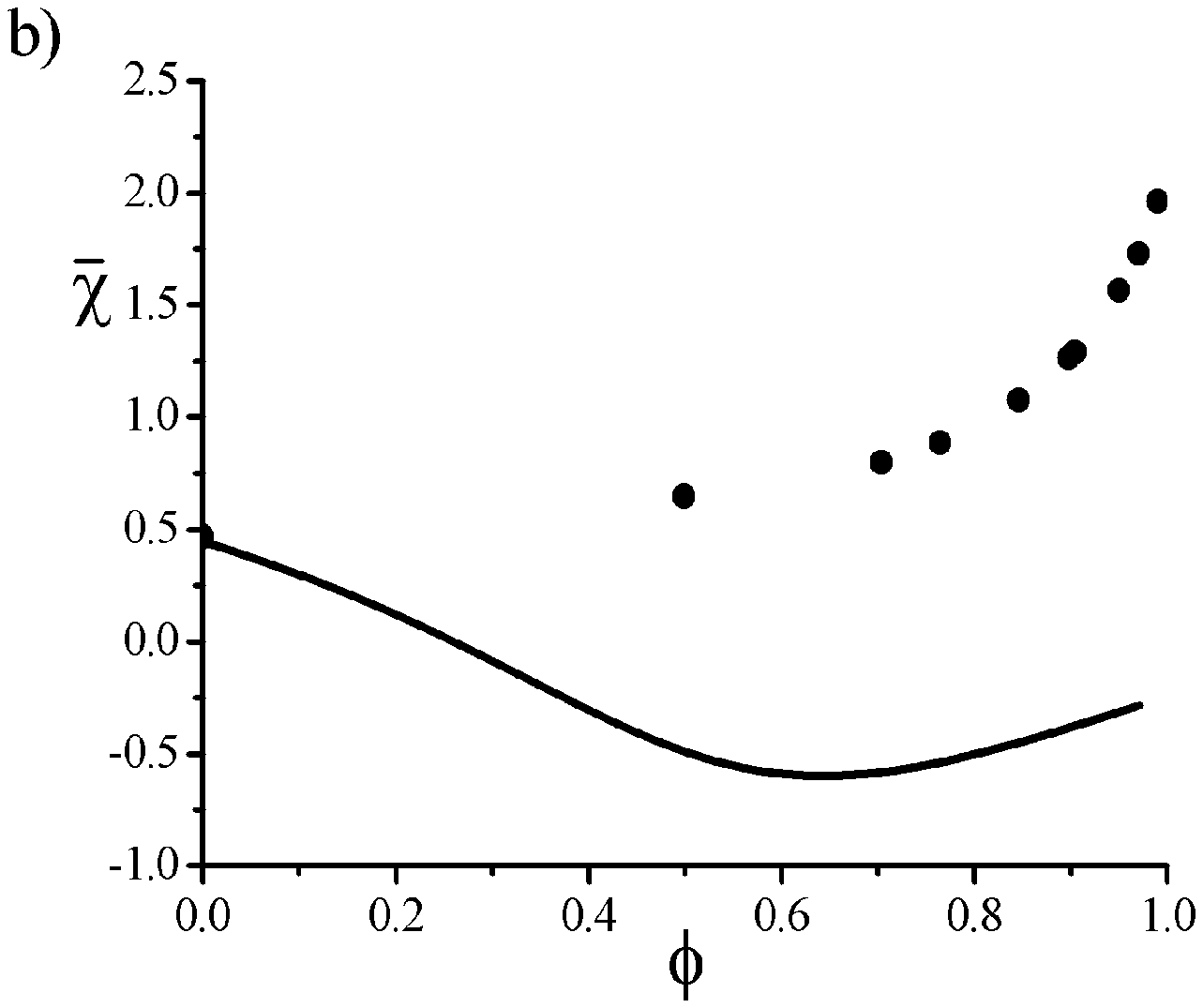}

Figure 5. $\overline{\chi}_{MB}$ {\em vs.} $\phi$ (continuous
line) calculated using $\chi_{AS}=$ $\chi_{BS}=\chi
=2.885-0.0036T$, $\chi _{AB}=0$ and $\Delta \epsilon
=6.38-2408/T$, the parameters used to fit the phase diagram of
PEO, qualitatively differ from the experimental values of
$\overline{\chi}$ (full circles) as measured with $N=3000$ at (a)
$T=328K$ and (b) $T=338K$. \vspace{-0.5cm}
\end{figure}

However, in the MB model $P_{e}$ depends on $\phi$ even when
$K_{MB}$ does not because of the loss of translational entropy of
the bound water. The stability condition for the
$B+S\rightleftarrows A$ reaction, $\partial^{2}F_{int}/\partial
P^{2}>0,$\ leads to

\begin{equation}
\frac{\partial ^{2}F_{int}}{\partial P^{2}}=\phi \left[
\frac{1}{P(1-P)}-2\phi \chi_{AB}+\frac{\phi }{1-\phi-P\phi}\right]
>0. \label{V5}
\end{equation}

Using $\overline{\chi}=-\frac{\partial}{\partial \phi
}\frac{F_{int}}{kT\phi}$ and the equilibrium condition (\ref{V4})
we find
\begin{eqnarray}
\overline{\chi}_{MB}&=&P_{e}\chi _{AS}+(1-P_{e})\chi _{BS}-\chi
_{AB}P_{e}(1-P_{e})+\nonumber \\
&&\frac{P_{e}}{\phi }+\frac{1}{\phi^{2}}\ln \left(
1-\frac{P_{e}\phi }{1-\phi }\right).  \label{V6}
\end{eqnarray}
While the first three terms are identical in form to
$\overline{\chi }_{K}$ these are now supplemented by two
additional terms arising from the translational entropy of the
water. When $\chi _{AS}=\chi _{BS}=\chi $ and $\chi_{AB}=0$ the
interaction terms are constant, $\overline{\chi }_{K}=\chi$, and
the $\phi $ dependence of $\overline{\chi }_{MB}$ reflects solely
the entropic contribution. Again we obtain $\partial P_{e}(\phi
)/\partial \phi $ by differentiating the equilibrium condition,
(\ref{V4}), with respect to $\phi $ finding
\begin{equation}
\frac{\partial P_{e}(\phi )}{\partial \phi }=\frac{\partial
\overline{\chi}_{MB}}{\partial P_{e}}\frac{\phi }{\partial
^{2}F_{int}/\partial P^{2}}. \label{V7}
\end{equation}

In equation (\ref{V7}), and later in (\ref{V8}), $\partial
^{2}F_{int}/\partial P^{2}$ denotes the value of (\ref{V5}) in
equilibrium. While equation (\ref{V7}) is similar in form to
(\ref{IV7}) the two are not identical because the two models yield
different $P_{e}$ and because $\partial \overline{\chi }_{MB}/\partial
 P_{e}\neq \partial \overline{\chi }_{K}/\partial P_{e}$.
 As before the sign of $\partial P_{e}(\phi )/\partial
\phi$ is determined by $\partial \overline{\chi }_{MB}/\partial
P_{e}$. This reflects the concentration dependence of
$K_{MB}^{^{\prime }}=(1-\phi-P\phi)K_{MB}$ in particular,
$\partial \overline{\chi }_{MB}/\partial P_{e}=\partial \ln
K_{MB}^{^{\prime }}/\partial \phi$.\ Upon substituting $
\partial P_{e}(\phi )/\partial \phi $ into $\partial \overline{\chi }
_{MB}/\partial \phi $ as found from (\ref{V6}) we obtain \
\begin{eqnarray}
\frac{\partial \overline{\chi }_{MB}}{\partial \phi }&=&\left(
\frac{\partial \overline{\chi }_{MB}}{\partial
P_{e}}\right)^{2}\frac{\phi }{\partial ^{2}F_{int}/\partial
P^{2}}- \nonumber \\
&&\frac{2}{\phi ^{3}}\ln \left( 1-\frac{P_{e}\phi }{1-\phi
}\right)- \nonumber \\
&&\frac{P_{e}}{\phi ^{2}}\left[ 1+\frac{1}{(1-\phi )(1-P_{e}\phi
-\phi )}\right]  \label{V8}
\end{eqnarray}

The first two terms are positive. Of these, the first is analogous
to the corresponding result within the K model. The third term is
negative. As a result, $\partial \overline{\chi}_{MB}/\partial
\phi$ is no longer positive definite. Depending on the choice of
$\chi_{AS}$, $\chi_{BS}$, $\chi_{AB}$ and $\Delta \epsilon$ the
calculated $\overline{\chi }_{MB}$ can be monotonically increasing
or decreasing as well as exhibit a minimum or a maximum (Figure
4). In the case of PEO the calculated $\overline{\chi}_{MB}(\phi
)$, utilizing the parameters used to fit the phase diagram of PEO,
differs qualitatively from the experimentally measured
$\overline{\chi} (\phi)$ (Figure 5).

\section{Discussion}

As we have seen, two versions of the ``hydrophobic-hydrophilic''
two-state model for water-soluble polymers in aqueous solutions
lead to a $\phi $ dependent $\chi _{eff}$. The K-version, where
the two states undergo unimolecular intrachain conversion, results
in $\partial \overline{\chi } /\partial \phi >0$. Using
$\chi_{AS}$, $\chi _{BS}$, $\chi_{AB}$ and $\Delta \epsilon$ as
obtained by fitting the phase diagram of PEO this model yields
$\overline{\chi}$ that agrees semiquantitatively with the
experimentally observed values\cite{Molyneux,Rowlinson} (Figure
3). Within the MB model the hydrophilic monomeric state binds a
water molecule. As a result, the interconversion reaction is
bimolecular and the translational entropy of the water plays a
role in determining the equilibrium state. In this model
$\overline{\chi}$ can display a number of scenarios (Figure 4):
$\overline{\chi}(\phi)$ can be monotonically increasing,
monotonically decreasing or exhibit an extremum (maximum or
minimum).

The $\overline{\chi}$ {\em vs}. $\phi$ as calculated with the
parameters used to fit the phase diagram of PEO, differs
qualitatively from the experimentally obtained curve(Figure 5).
However, the $\partial \overline{\chi }/\partial \phi <0$ behavior
allowed by this model is of interest since it has been observed in
aqueous solutions of the neutral water-soluble polymer
PVP\cite{Martien} (Figure 2). These results stress the importance
of using the experimental values of $\overline{\chi}$ in fitting
$\chi _{AS}$, $\chi_{BS}$, $\chi_{AB}$ and $\Delta \epsilon$ and
in evaluating the performance of the model. From the perspective
of the general theory of polymers it is of interest that the
``two-state'' models can account for the dependence of
$\chi_{eff}$ on $\phi$, $T$ and the pressure while retaining the
principal approximations of the Flory-Huggins theory {\em i.e.,}
incompressibility, monomer and solvent of identical size and shape
and local composition that equals the global one.

At this point it is helpful to consider alternative mechanisms
leading to $\chi_{eff}(\phi)$. For brevity we limit this
discussion to two physically transparent models. One is the
$n$-cluster model proposed by de Gennes for aqueous solutions of
PEO.\cite{n-Cluster} This model was motivated by two experimental
observations concerning the behavior of such solutions: (i) the
interpretation of calorimetric measurements in terms of the Flory
mixing free energy yields a $\overline{\chi}$ that increases
strongly with $\phi$,\cite{Molyneux} and (ii) reports of formation
of aggregates in concentrated solutions of PEO.\cite{Burchard}
Within this model the concentration dependence of $\chi_{eff}$ is
attributed to attractive interactions leading to stable clusters
of $n>2$ monomers while binary monomer-monomer interactions remain
repulsive. This is another variation of a two-state model
involving a dynamic equilibrium between ``clustered'' and
``unclustered'' monomers. In molecular terms, a $n$-cluster may
correspond to a micelle or a mixed helix. The formation of the
$n$-clusters gives rise to an additional term, $-\rho(T)\phi^{n}$
(with $\rho >0$), in the interaction free energy. Thus, the mixing
free energy per site is
\begin{eqnarray}
\frac{F_{mix}}{kT}&=&\chi \phi (1-\phi )+\rho (T)(\phi
-\phi ^{n})+ \nonumber \\
&&\frac{\phi }{N}\ln \phi +(1-\phi )\ln (1-\phi )  \label{VI1}
\end{eqnarray}
leading to
\begin{equation}
\chi _{eff}=\chi +\rho \frac{1-\phi ^{n-1}}{1-\phi }.  \label{VI2}
\end{equation}
The corresponding $\overline{\chi}$ is
\begin{equation}
\overline{\chi }=\chi +\rho (n-1)\phi^{n-2}  \label{VI3}
\end{equation}
and $\partial \overline{\chi }/\partial \phi >0$. The $n$-cluster
model is indeed capable of rationalizing the two experimental
observations noted above. However of these, the second observation
is now a subject of debate.\cite{antiB} Furthermore the molecular
structure of the ethylenoxide monomers does not reveal amphiphilic
motifs. \ It is thus difficult to justify the assumption of
cluster formation in solutions of PEO. Accordingly, the validity
of this model in the case of aqueous solutions of neutral water
soluble polymers such as PEO is not obvious. On the other hand,
the model is indeed applicable to solutions of polysoaps where
formation of inter and intrachain micelles does
occur\cite{Polysoaps}. The second model, advanced by Painter {\em
et al}, is applicable to all polymer solutions.\cite{Painter}
Within this model, the $\phi$ dependence of $\chi_{eff}$ is
attributed to the interplay of intrachain and interchain contacts.
The authors argue that there is a probability $\gamma$ for
intrachain monomer-monomer contacts. As a rough approximation,
$\gamma $ can be identified with the monomeric volume fraction,
$\phi _{G}$, within a Gaussian coil of radius $R_{G}\approx
N^{1/2}a$ where $N$ is the polymerization degree and $a$ is the
monomer size, $\phi _{G}\approx Na^{3}/R_{0}^{3}\approx N^{-1/2}$.
In a lattice comprising of $N_{T}$ sites of coordination number
$z$ there are $N_{T}\phi $ sites occupied by monomers with a total
of $N_{T}z\phi $ adjacent sites. Of the $N_{T}z\phi$ adjacent
sites $N_{T}z\phi \gamma$ are occupied by monomers because of
intrachain contacts. The total number of free, unblocked adjacent
sites is thus $N_{T}z(1-\phi \gamma)$ while the number of free
sites adjacent to monomers is $N_{T}z\phi (1-\gamma)$. The
probability of a free site adjacent to a monomer is thus $\phi
(1-\gamma )/(1-\phi \gamma ).$ Accordingly, the number of
monomer-solvent contacts is $N_{T}(1-\phi )\phi (1-\gamma
)/(1-\phi \gamma )$. This expression allows for the requirement
that a solvent molecule occupies an unblocked site adjacent to a
monomer. Accordingly, the mixing energy term per site is
$E_{mix}/kT=\chi (1-\phi )\phi (1-\gamma )/(1-\phi \gamma)$
yielding an enthalpic contribution to $\chi _{eff}$
\begin{equation}
\chi _{eff}=\chi \frac{1-\gamma }{1-\phi \gamma}  \label{VI4}
\end{equation}
and to a corresponding enthalpic $\overline{\chi}$
\begin{equation}
\overline{\chi }=\chi \left( \frac{1-\gamma }{1-\phi \gamma }\right)^{2}
\label{VI5}
\end{equation}
where, again, $\partial \overline{\chi}/\partial \phi >0.$ In
their original paper, Painter {\em et al} supplemented this
enthalpic $\overline{\chi }$ by an entropic one, allowing for the
effect of the chain bending back on itself following the analysis
of Huggins.\cite{Huggins}

Aside from physical insight regarding the molecular origins of
$\chi_{eff}(\phi)$, the above discussion identifies certain
difficulties. Overall, it seems unlikely that one of the four
models described above will emerge fully victorious. By
construction, the $n$-cluster model is applicable only to
solutions of associating polymers. The two ``two-state'' models
are suitable candidates for the description of aqueous solutions
of neutral water-soluble polymers exhibiting insolubility gap.
While the model of Painter {\em et al} applies, in principle, to
all polymeric systems it can not account for systems exhibiting
$\partial \overline{\chi }/\partial \phi <0$. Such behavior was
actually observed in both aqueous and non-aqueous solutions, yet
of the four models considered above only the MB model yields a
scenario involving $\partial \overline{\chi}/\partial \phi<0$.
Altogether one can thus envision situations where {\em all the
different mechanisms described may contribute simultaneously}.
Another discouraging observation concerns the number of parameters
involved. All four models introduce additional parameters that do
not appear in the familiar Flory-Huggins theory: $\gamma$ in the
model of Painter {\em et al}, $\rho $ and $n$ in the $n$-cluster
model, $\Delta \epsilon $, $\chi _{AB}$, $\chi_{AS}$ and $\chi
_{BS}$ for the two ``two-state'' models. The necessity to
unambiguously determine the additional parameters limits the
predictive power of the models. Clearly, this problem is even more
serious when a number of mechanisms contribute simultaneously to
the $\phi$ dependence of $\chi _{eff}$. Note however that this
last difficulty can be partially resolved in certain cases. Thus
the contribution of the mechanism of Painter {\em et al} can be
separated from the one due to the ``two-state'' mechanism.\ This
is because the mechanism of Painter {\em et al }is inherently a
polymeric effect that disappears in the monomeric limit, $N=1$. On
the other hand, {\em the two-state models also apply to solution
of the unpolymerized monomers} {\em i.e.,} $\chi_{eff}$ and
$\overline{\chi }$ are independent of $N$.

Following our discussion of alternative models for
$\chi_{eff}(\phi )$ it is helpful to summarize the evidence
favoring the two-state models. Using these models it was possible
to derive the phase diagram of PEO\cite{K,BBP} and its pressure
dependence.\cite{BBP} In addition, the K model was successfully
utilized to rationalize a variety of systems involving PEO,
Polypropyleneoxyde (PPO)and PEO-PPO copolymers.\cite{swedes} These
successes lend support to the validity of the two-state model. On
the negative side, there is relatively little direct evidence for
the existence of the two monomeric states. Computational studies
lend the most direct support to the two-state models. The K model
is supported by recent molecular dynamics study that monitored the
population of hydrophilic and hydrophobic conformations as a
function of temperature and concentration and relates their
relative stability to polar interactions.\cite{Smith} This study
utilized force fields obtained from quantum mechanical
calculations of dimethoxyethane.\cite{Smith1} Somewhat different
conclusions were reached in studies of higher oligo(ethylene
oxides).\cite{WKG} In particular, these studies (i) support the MB
model (ii) stress the importance of conformers allowing formation
of ``water bridges'' {\em i.e.}, where a single water molecule
forms two hydrogen bonds with the oxygens in the oligo(ethylene
oxides). One such conformer, though not the most stable, is
helical. (iii) allow to rationalize force measurements of
PEO\cite{ORG} and, in particular, the difference in the extension
force laws found in water and in non- aqueous solvent.\cite{KWG}
Thus while a consensus on the computational results is yet to
emerge, this approach may eventually allow to identify the
relevant monomeric states and to obtain $\Delta \epsilon$,
$\chi_{AB}$, $\chi_{AS}$ and $\chi_{BS}$.


\begin{thebibliography}{99}
\bibitem{nom}  In the literature,\cite{wolf} $\chi_{eff}$ is sometimes
denoted by $g$ while $\chi $ denotes $\overline{\chi }=\chi
_{eff}-(1-\phi )\partial \chi _{eff}/\partial \phi$. Our notation
aimes to avoid confusion with the customary usage of $\chi$.
\end{thebibliography}
\end{document}